# About the Abraham force in a conducting medium

Yurii A. Spirichev

The State Atomic Energy Corporation ROSATOM, "Research and Design Institute of Radio-Electronic Engineering" - branch of Federal Scientific-Production Center "Production Association "Start" named after Michael V.Protsenko",

Zarechny, Penza region, Russia

E-mail: yurii.spirichev@mail.ru

(Dated: May 20, 2017)

**Abstract**

The article is devoted to the Abragam force in a conducting medium. From the energy-momentum tensor of the electromagnetic field interaction with moving charges the equation for the Abraham force is obtained. It is shown that the Abraham force in a conducting medium, as well as in a dielectric medium, has vortical character. Current density vectors and vector potential must be non-colinear.

**Keywords**: the energy-momentum tensor, the electromagnetic forces, electromagnetic momentum, the Abraham force.

**Content:**
1. Introduction
2. Electromagnetic energy-momentum tensor of the electromagnetic field interaction with moving charges
3. Mechanical energy-momentum tensor
4. Equations of energy conservation laws and momentum of the interaction of an electromagnetic field with moving charges
5. The Abraham force in a cconducting medium
6. Conclusion
   References

## 1. Introduction

Interaction of the electromagnetic field with a medium is fundamental for many physical phenomena and is of great practical importance. One of these phenomena is the Abraham electromagnetic force. A number of recent works have been devoted to the interaction of the electromagnetic field with a medium [1 - 15]. In general, the Abraham force is written in the form of the difference of the time derivatives of the electromagnetic momentum in Minkowski and Abraham forms.

In the article [9] it is shown that the equation for the Abraham force in a dielectric medium follows from the Minkowski energy-momentum tensor. A new energy-momentum tensor for a dielectric medium was obtained in [6], from which follows the equation for the Abraham force in the form:

$$\mathbf{F}_A = \partial_t \mathbf{g}^M - \partial_t \mathbf{g}^A = \frac{1}{4\pi \cdot c} \partial_t (\mathbf{D} \times \mathbf{B} - \mathbf{E} \times \mathbf{H}) = \frac{1}{4\pi} \nabla \times (\mathbf{E} \times \mathbf{D} + \mathbf{B} \times \mathbf{H}) \quad (1)$$

Where $\mathbf{g}^M$, $\mathbf{g}^A$, are electromagnetic momentum densities, in Minkowski and Abraham forms respectively; $\mathbf{E}$, $\mathbf{H}$, $\mathbf{D}$, $\mathbf{B}$ - intensities of the electromagnetic field and electromagnetic induction. It follows from the equation that the Abraham force is a vortex force and it arises when the vectors $\mathbf{D}$ and $\mathbf{E}$, $\mathbf{H}$ and $\mathbf{B}$ vary in time and are non-collinear.

In equation (1), the macroscopic model of a dielectric medium is described by the vector of electric induction $\mathbf{D}$ and the magnetic permeability $\mathbf{H}$, which usually include dielectric and magnetic permeabilities of a medium. It may be naturally suggested that the Abraham force exists for a microscopic model of a medium with electric charges and currents. The Abraham force equation for microscopic model of a conducting medium does not exist in modern electrodynamics. The Abraham force equation for a conducting medium, as well as equation (1), must follow from the energy-momentum tensor of the interaction of the electromagnetic field with charges and currents.

The aim of this paper is to obtain the Abraham force equation for a conducting medium.

Space-time geometry is taken to be in the form of pseudo-Euclidean Minkowski space (ct, ix, iy, iz). The electromagnetic potential is taken to be in the form $\mathbf{A}_\mu$ ($\varphi / c$, $i\mathbf{A}$), where $\varphi$ and $\mathbf{A}$ are the scalar and vector potentials of the electromagnetic field. Four-dimensional charges velocity is taken to be in the form $\mathbf{J}_\mu (c \cdot \rho, i \cdot \mathbf{J})$, where $\rho$ and $\mathbf{J}$ are electric charges density and a three-dimensional current density vector respectively. Four-dimensional velocity of charges is taken to be in the form $\mathbf{V}_\mu (c, i \cdot \mathbf{V})$, where $\mathbf{V}$ is a three-dimensional charge velocity vector. The four-dimensional density of the mechanical momentum is taken to be in the form $\mathbf{P}_\nu (c \cdot m, i \cdot \mathbf{p})$, where $\mathbf{m}$ and $\mathbf{p}$ is mass density and three-dimensional mechanical momentum respectively.

## 2. Electromagnetic energy-momentum tensor of the electromagnetic field interaction with moving charges

Electromagnetic energy-momentum tensor of the interaction of the electromagnetic field with charges and currents $T^E_{\mu\nu}$ is obtained as the tensor product of the four-dimensional vector potential of the electromagnetic field $\mathbf{A}_\mu$ and the four-dimensional current density $\mathbf{J}_\nu$:

$$T_{\mu\nu}^E = \mathbf{A}_\mu \otimes \mathbf{J}_\nu = \begin{pmatrix} \rho\cdot\varphi & \frac{1}{c}i\cdot\varphi\cdot J_x & \frac{1}{c}i\cdot\varphi\cdot J_y & \frac{1}{c}i\cdot\varphi\cdot J_z \\ i\cdot c\cdot\rho\cdot A_x & -A_x\cdot J_x & -A_x\cdot J_y & -A_x\cdot J_z \\ i\cdot c\cdot\rho\cdot A_y & -A_y\cdot J_x & -A_y\cdot J_y & -A_y\cdot J_z \\ i\cdot c\cdot\rho\cdot A_z & -A_z\cdot J_x & -A_z\cdot J_y & -A_z\cdot J_z \end{pmatrix} = \begin{bmatrix} W & \frac{1}{c}i\cdot\mathbf{S} \\ i\cdot c\cdot\mathbf{p} & t_{mn} \end{bmatrix} \quad (2)$$

Where $W = \rho\cdot\varphi$ - total electromagnetic energy density, $\mathbf{g} = \rho\cdot\mathbf{A}$ is the density of the electromagnetic momentum; $\mathbf{S} = \varphi\cdot\mathbf{J}$ - electromagnetic energy flux density; $t_{mn} = -\mathbf{A}_m \otimes \mathbf{J}_n$ - three-dimensional tensor of electromagnetic momentum flux density or stress tensor. The tensor (2) corresponds to the canonical energy-momentum tensor [16]. From it follows the linear invariant of electromagnetic energy density:

$$I = \varphi\cdot\rho - \mathbf{A}\cdot\mathbf{J}$$

This invariant is known in electrodynamics as a generalized energy density of electromagnetic interaction and represents Lagrange function with the opposite sign for charges moving in the electromagnetic field with density ρ.

From tensor $T_{\mu\nu}^E$, in the form of its four-dimensional divergences (from now on there is no necessity to draw a distinction between covariant and contravariant indices), the equations of energy-momentum conservation laws follow:

$$\partial_\mu T_{\mu\nu}^E = \partial_\mu T_{\mu\nu}^M \quad \text{и} \quad \partial_\nu T_{\mu\nu}^E = \partial_\nu T_{\mu\nu}^M$$

Where $T_{\mu\nu}^M$ is the mechanical energy-momentum tensor.

## 2. Mechanical energy-momentum tensor

We will obtain mechanical energy-momentum tensor as the tensor product of the four-dimensional charges velocity vector $\mathbf{V}_\mu(c, i\cdot\mathbf{V})$ and the four-dimensional vector of momentum density $\mathbf{P}_\nu(c\cdot m, i\cdot\mathbf{p})$:

$$T_{\mu\nu}^M = \mathbf{V}_\mu \otimes \mathbf{P}_\nu = \begin{pmatrix} m\cdot c^2 & i\cdot c\cdot p_x & i\cdot c\cdot p_y & i\cdot c\cdot p_z \\ i\cdot c\cdot p_x & -V_x\cdot p_x & 0 & 0 \\ i\cdot c\cdot p_y & 0 & -V_y\cdot p_y & 0 \\ i\cdot c\cdot p_z & 0 & 0 & -V_z\cdot p_z \end{pmatrix} = \begin{bmatrix} W & \frac{1}{c}i\cdot\mathbf{S} \\ i\cdot c\cdot\mathbf{p} & t_{mn}^M \end{bmatrix} \quad (3)$$

Where $W = m\cdot c^2$ - density of total energy per unit volume of the medium; $\mathbf{S} = \mathbf{p}\cdot c^2$ - Energy flux density vector. Three-dimensional tensor of mechanical momentum flux density $t_{mn}^M$ or a stress tensor can be considered as tensor of the kinetic energy density:

$$t_{ik}^M = \begin{bmatrix} -p_x\cdot V_x & 0 & 0 \\ 0 & -p_y\cdot V_y & 0 \\ 0 & 0 & -p_z\cdot V_z \end{bmatrix} = -\mathbf{p}\cdot\mathbf{V}$$

Its mixed components equal zero, since the scalar products of the mixed velocity components are zero. The mechanical energy-momentum tensor (3) is symmetric. Its trace is a four-dimensional invariant:

$$Tr = m \cdot c^2 - \mathbf{p} \cdot \mathbf{V}$$

## 4. Equations of energy conservation laws and momentum of the interaction of an electromagnetic field with moving charges

From tensors $T^E_{\mu\nu}$ and $T^M_{\mu\nu}$ follow the equations of energy and momentum conservation laws in the form of their four-dimensional divergences $\partial_\mu T^E_{\mu\nu} = \partial_\mu T^M_{\mu\nu}$ and $\partial_\nu T^E_{\mu\nu} = \partial_\nu T^M_{\mu\nu}$. Let's write them in expanded form:

$$\frac{1}{c^2}\partial_t(\varphi \cdot \rho) + \nabla(\rho \cdot \mathbf{A}) = \partial_t m + \nabla \cdot \mathbf{p} \tag{4}$$

$$\frac{1}{c^2}\partial_t(\varphi \cdot \mathbf{J}) + \partial_m(\mathbf{A}_m \otimes \mathbf{J}_n) = \partial_t \mathbf{p} + \nabla(\mathbf{p} \cdot \mathbf{V}) \tag{5}$$

$$\frac{1}{c^2}\partial_t(\varphi \cdot \rho) + \frac{1}{c^2}\nabla(\varphi \cdot \mathbf{J}) = \partial_t m + \nabla \cdot \mathbf{p} \tag{6}$$

$$\partial_t(\rho \cdot \mathbf{A}) + \partial_n(\mathbf{A}_m \otimes \mathbf{J}_n) = \partial_t \mathbf{p} + \nabla(\mathbf{p} \cdot \mathbf{V}) \tag{7}$$

Eq. (4) and (6) are energy conservation equations, and Eq. (5) and (7) are conservation equations for two forms of electromagnetic momentum density. These forms of the electromagnetic momentum density are similar to the forms of the Minkowski and Abraham. Eq. (4) and (6) are followed by the equation:

$$\nabla(\rho \cdot \mathbf{A}) = \frac{1}{c^2}\nabla(\varphi \cdot \mathbf{J}) \tag{8}$$

The divergences of the two forms of the electromagnetic momentum density are equal. Eq. (5) and (7) are followed by the equation:

$$\frac{1}{c^2}\partial_t(\varphi \cdot \mathbf{J}) + \partial_m(\mathbf{A}_m \otimes \mathbf{J}_n) = \partial_t(\rho \cdot \mathbf{A}) + \partial_n(\mathbf{A}_m \otimes \mathbf{J}_n) \tag{9}$$

## 5. The Abraham force in a conducting medium

Eq. (5) and (7) can be considered as equations of balance of the density of electromagnetic and mechanical forces. Eq. (5) and (7) are conservation equations for two forms of electromagnetic momentum density. These forms of electromagnetic momentum density are analogues of two forms of the momentum density in a dielectric medium: the Minkowski form and the Abraham form. The Abraham force density is equal to the difference of time derivatives of these two forms of electromagnetic momentum density (1). To obtain the equation for the Abragam force density, we write the Eq. (9) in the form:

$$\partial_t(\rho \cdot \mathbf{A}) - \frac{1}{c^2}\partial_t(\varphi \cdot \mathbf{J}) = \partial_m(\mathbf{A}_m \otimes \mathbf{J}_n) - \partial_n(\mathbf{A}_m \otimes \mathbf{J}_n)$$

The left-hand side of this equation is the difference of time derivatives of the two forms of the electromagnetic momentum; similar to the Eq. (1), i.e., it is the Abraham force density in a conducting medium. Expanding the right-hand side of the equation, we will obtain the equation for the Abraham force in the form:

$$\mathbf{F}_A = \partial_t(\rho \cdot \mathbf{A}) - \frac{1}{c^2}\partial_t(\varphi \cdot \mathbf{J}) = \nabla \times (\mathbf{A} \times \mathbf{J}) \qquad (10)$$

The equation for the Abraham force in conducting medium has the form which is similar to the Eq. (1) of this force in a dielectric medium. It follows from Eq. (10) that the Abraham force in a conducting medium, like in the Eq. (1), is a vortex force, while the vectors **A** and **J** must be non-collinear.

## 6. Conclusion

The Abraham force exists not only in a dielectric medium, but also in a conducting medium. Its equation follows from the electromagnetic energy-momentum tensor of interaction of electromagnetic field with charges and currents. The Abraham force in a conducting medium, as well as in a dielectric medium, is a vortex force, while the vectors **A** and **J** must be non-collinear.

**References**
1. I. Brevik, *Minkowski momentum resulting from a vacuum–medium mapping procedure, and a brief review of Minkowski momentum experiments*, Annals of Physics 377 (2017) 10–21.
2. Rodrigo Medina and J. Stephany, *The energy-momentum tensor of electromagnetic fields in matter*, arXiv:1703.02109.
3. Mario G. Silveirinha, *Revisiting the Abraham-Minkowski Dilemma*, arXiv: 1702.05919.
4. Joseph J. Bisognano, *Electromagnetic Momentum in a Dielectric: a Back to Basics Analysis of the Minkowski-Abraham Debate*, arXiv: 1701.08683.
5. Yurii A. Spirichev, *Equation for the Abraham force in non-conducting medium and methods for its measurement*, arXiv: 1704.03368.
6. Yurii A. Spirichev, *A new form of the energy-momentum tensor of the interaction of an electromagnetic field with a non-conducting medium. The wave equations. The electromagnetic forces*, arXiv: 1704.03815.
7. Michael E. Crenshaw, *The Role of Conservation Principles in the Abraham--Minkowski Controversy*, arXiv: 1604.01801.

8. C.J. Sheppard, B.A. Kemp, Phys. Rev. A 93 (2016) 053832.
9. V.V. Nesterenko, A.V. Nesterenko, J. Math. Phys. 57 (2016) 092902.
10. C. Wang, "Is the Abraham electromagnetic force physical?" Optik, (2016) 127, 2887–2889.
11. Pablo L. Saldanha, J. S. Oliveira Filho, *Hidden momentum and the Abraham-Minkowski debate*, arXiv: 1610.05785.
12. Massimo Testa, *A Comparison between Abraham and Minkowski Momenta*, Journal of Modern Physics, 2016, 7, 320-328.
13. L. Zhang, W. She, N. Peng, U. Leonhardt, New J. Phys. 17 (2015) 053035
14. G. Verma and K. P. Singh, Phys. Rev. Lett. 115, 143902 (2015)
15. Choi H, Park M, Elliott D S, Oh K, *Optomechanical Measurement of the Abraham Force in an Adiabatic Liquid Core Optical Fiber Waveguide*, arXiv: 1501.05225.
16. Landau L D, Lifshits E M *The Classical Theory of Fields* (Oxford: Pergamon Press, 1983).